\documentclass[twocolumn]{aastex63}

\newcounter{daggerfootnote}

\newcommand{\mgii}{Mg\,\textsc{ii}}
\newcommand{\oiii}{O\,\textsc{iii}}

\newcommand{\nii}{N\,\textsc{ii}}
\newcommand{\sii}{S\,\textsc{ii}}
\newcommand{\neiii}{Ne\,\textsc{iii}}

\newcommand{\srcname}{GN-72127}
\newcommand{\zsource}{4.1288}

\newcommand{\jwst}{\textit{JWST}}

\definecolor{mycol}{rgb}{0,0,1}

\received{n/a}
\revised{n/a}
\accepted{n/a}

\submitjournal{\apj}

\shorttitle{AGN Feedback and a Quenched LRD}
\shortauthors{Kokorev et al.}

\usepackage{xcolor}
\usepackage{longtable}
\usepackage{lipsum}
\usepackage{amsmath}
\usepackage[normalem]{ulem}
\usepackage[para]{threeparttable}
\usepackage{enumitem}
\usepackage[encapsulated]{CJK}

\begin{document}

\title{Silencing the Giant: Evidence of AGN Feedback and Quenching in a Little Red Dot at $z=4.13$}

\correspondingauthor{Vasily Kokorev}
\email{vkokorev@utexas.edu}

\author[0000-0002-5588-9156]{Vasily Kokorev}
\affiliation{Department of Astronomy, The University of Texas at Austin, Austin, TX 78712, USA}

\author[0000-0002-0302-2577]{John Chisholm}
\affiliation{Department of Astronomy, The University of Texas at Austin, Austin, TX 78712, USA}

\author[0000-0003-4564-2771]{Ryan Endsley}
\affiliation{Department of Astronomy, The University of Texas at Austin, Austin, TX 78712, USA}

\author[0000-0001-8519-1130]{Steven L. Finkelstein}
\affiliation{Department of Astronomy, The University of Texas at Austin, Austin, TX 78712, USA}

\author[0000-0002-5612-3427]{Jenny E. Greene}
\affiliation{Department of Astrophysical Sciences, Princeton University, 4 Ivy Lane, Princeton, NJ 08544}


\author[0000-0003-3596-8794]{Hollis B. Akins}
\affiliation{Department of Astronomy, The University of Texas at Austin, Austin, TX 78712, USA}

\author[0000-0003-0212-2979]{Volker Bromm}
\affiliation{Department of Astronomy, The University of Texas at Austin, Austin, TX 78712, USA}

\author[0000-0002-0930-6466]{Caitlin M. Casey}
\affiliation{Department of Astronomy, The University of Texas at Austin, Austin, TX 78712, USA}

\author[0000-0001-7201-5066]{Seiji Fujimoto}
\affiliation{Department of Astronomy, The University of Texas at Austin, Austin, TX 78712, USA}
\affiliation{Cosmic Dawn Center (DAWN), Niels Bohr Institute, University of Copenhagen, Jagtvej 128, K{\o}benhavn N, DK-2200, Denmark}

\author[0000-0002-2057-5376]{Ivo Labb\'e}
\affiliation{Centre for Astrophysics and Supercomputing, Swinburne University of Technology, Melbourne, VIC 3122, Australia}

\author[0000-0003-2366-8858]{Rebecca L. Larson}
\affiliation{School of Physics and Astronomy, Rochester Institute of Technology, 84 Lomb Memorial Drive, Rochester, NY 14623, USA}

\begin{abstract}
The \textit{James Webb Space Telescope} (\textit{JWST}) has uncovered a ubiquitous population of dust-obscured compact sources at $z\gtrsim 4$. Many of these objects exhibit signs of active galactic nucleus (AGN) activity, making their study crucial for understanding the formation of supermassive black holes (SMBHs) and their growth with host galaxies. In this work, we examine low and medium resolution \textit{JWST}/NIRSpec spectra from the JADES GTO public data release in the GOODS-N field of a red, luminous ($M_{\rm B}\sim-22.2$ mag) and compact ($<500$ pc) source at $z=4.13$.
The rest-optical ($\lambda_{\rm rest} > 4000$ \AA) continuum of this source is strongly dominated by a massive (log$_{10}$[$M_*/M_\odot] \sim 10.6$), quenched (log$_{10}$[sSFR/yr$^{-1}$] $< -11$) galaxy, as indicated by the clear presence of a Balmer break and stellar absorption lines. Star-formation history modeling reveals a starburst episode followed by rapid quenching about 200 Myr ago. The spectrum shows extremely broad (FWHM $\sim 2500$ km/s) H$\alpha$ emission and elevated optical line ratios, indicating an actively accreting SMBH. Moreover, our work has potentially revealed clear AGN signatures in the rest-UV in LRDs for the first time, via a detection of a strong Ly$\alpha$ emission and a broad \mgii\, doublet. The derived black hole mass of log$_{10}$($M_{\rm BH}/M_\odot) \sim 7.3$ results in $M_{\rm BH}/M_*\sim 0.04$ \%, consistent with the local relations, unlike the elevated ratios in other high-$z$ reddened AGN. Finally, we use \textit{JWST} data from AGN at $z=4-10$ to explore an evolutionary link between high-$z$ reddened AGN, early quiescent galaxies, and local ellipticals.
\end{abstract}


\section{Introduction} \label{sec:intro}

Observations over the past few decades have established that super massive black holes (SMBHs) exist in the cores of nearly all massive galaxies locally \citep{kormendy97,richstone98,kormendy13}. The growth of SMBHs is driven by gas accretion, designating such black holes as an active galactic nucleus (AGN). 
Powerful AGN drive the energy and momentum in the interstellar medium (ISM) of galaxies, heating up the gas and slowing down star formation. As such the growth of stars in galaxies and the physics of their SMBHs are tightly linked \citep{heckman14}.
\par
The \textit{James Webb Space Telescope} (\textit{JWST})
is at the center stage of discovery of high-$z$ AGN, getting ever closer to their epoch of formation. The first two years of its operation have marked a discovery of numerous, previously-missing, UV-faint AGN, shifting our paradigm of SMBH formation \citep{smith2019}. These have been identified through a combination of directly observing broad lines
\citep[Type I;][]{goulding23,harikane23_agn,
juodzbalis24,kocevski23,larson23,maiolino23,maiolino23b,matthee23,ubler23}, inferring their AGN nature through ionization \citep[Type II;][]{chisholm24,scholtz23}, or some combination of color and morphology \citep{barro23,iani24,yang23}.
\par
Among these, an enigmatic population of compact red sources, so called ``little red dots'' \citep[LRDs;][]{matthee23}, stands out in particular. These show a characteristic double-break feature in their spectral energy distributions (SEDs), with seemingly dust-free UV emission and a steep red slope in the rest-optical \citep{furtak23,kocevski23,labbe23}.
These were initially speculated to be massive compact galaxies at early epochs \citep{barro23,labbe23_nat}, however spectroscopic follow-up observations in many of these sources have shown clear evidence of broad Balmer emission and high-ionization, all implicating them as actively accreting SMBHs \citep{fujimoto23_uncover,furtak24,greene24,killi23,kocevski23,kokorev23c,matthee23,ubler23}.
\par
Staggeringly, SMBHs in a large number of high-$z$ AGN, including LRDs, are overmassive,
with the ratio between their black hole and stellar mass ($M_{\rm BH}/M_*$) reaching upwards of $\sim 30 - 50$ \% \citep{bogdan23,goulding23,juodzbalis24,kokorev23c} which significantly deviates from the local relations \citep[e.g.][]{greene16} by more than a few orders of magnitude \citep{pacucci23,pacucci24}. Seemingly the growth of these black holes completely outpaces that of their host galaxies, at least at early times \citep[e.g.,][]{jeon2024}. In contrast, quite a few of these systems have been shown to be fully consistent with the local $M_{\rm BH}-\sigma$ relations \citep{maiolino23b}, which could point toward the presence of gas reservoirs that sustain the growth of the central AGN, but which are inefficient at forming stars.
\par
Detailed spectroscopic \citep{greene24,maiolino23b,matthee23} and photometric \citep{akins24, kocevski24,kokorev24} forays into their number densities reveal LRDs to be much more abundant compared to the UV-selected quasars at $z\sim4-6$. They appear to account for a significant fraction of all broad-line selected AGN \citep{harikane23_agn} and even for a few percent of the total galaxy population at $z>6$.
The over-abundance of AGN  with large BH masses relative to local scaling relations in the first few billion years of cosmic evolution can therefore truly test our models and preconceived vision of BH seeding as well as their evolution alongside their host galaxies \citep{dayal24,dayal24b,natarajan23}.
\par
What has yet remained unclear is the origin of the observed rest-UV emission in LRDs. So far, the interpretations range from scattered light from the AGN itself or an unobscured host galaxy component \citep[][]{akins23,barro23,greene24,kocevski23,kocevski24,labbe23}. Given the similarities between the observed UV slopes of quasars and star-forming galaxies, neither deep \textit{JWST} spectra \citep{greene24,kokorev23c}, nor vast photometric samples of LRDs \citep{akins24,kocevski24,kokorev24} have been able to make a conclusive determination. Lacking a complete understanding of the rest-UV side of LRDs prevents us from robustly deriving properties such as the stellar mass ($M_*$), star formation rate (SFR), dust attenuation ($A_{\rm V}$) and ionization state in these sources. This in turn limits our ability to attain an accurate view of the impact the AGN has on its host.
\par
Finally, despite hiding behind a thick veil of dust, these objects can be quite bright, sometimes reaching observed F444W magnitudes of $22-23$ (\citealt{akins24,kokorev24,wang24}; Labb\'e et al. in prep.), and yet a similar class of objects has not been detected at either intermediate ( $z\lesssim4$) redshifts or locally \citep{kocevski24}. Seemingly this complete absence of overmassive SMBHs at late times implies that such AGN only exist at early times and on short timescales, followed by an uncertain descendant population. Recent works by \citet{wang24} and Labb\'e et al. (in prep.) show that some reddened broad-line (BL) AGN already have quite evolved stellar populations by $z\sim4$, as implied by their strong Balmer breaks and absorption lines. Therefore, older LRDs may be positioned more in line with quiescent galaxies found at similar redshifts \citep{carnall23,de_graaff24}, rather than extreme AGN at high-$z$. Could this be the next step of overmassive SMBH evolution?
\par
In this work, we report the discovery of BL AGN emission in an otherwise stellar dominated galaxy at $z=4.13$. The deep \textit{JWST}/NIRSpec
Micro-Shutter Assembly (MSA) PRISM and medium resolution grating spectra in the GOODS-N field \citep{deugenio24}, allow us to identify the presence of a broad H$\alpha$ emission alongside an an array of [\oiii], [\nii] and [\sii] lines, all showing line ratios typical for AGN activity. At the same time, we also observe a strong Balmer break and a multitude of Balmer absorption lines indicative of an evolved stellar population. While this galaxy is completely consistent with the little red dot (LRD) color and compactness criteria, the observed red rest-optical color likely originates from an evolved stellar population, rather than from the reddened AGN continuum emission. Through the examination of black hole to host mass ratio we conclude that \srcname\, has more in common with quiescent galaxies \citep{carnall23,de_graaff24} and local ellipticals rather than the extreme high-$z$ AGN observed with \textit{JWST}. Using that information we attempt to draw an evolutionary sequence that connects objects with elevated $M_{\rm BH}/M_*$ at high-$z$ to the evolved galaxies with residual AGN activity we observe later on. Moreover, our work has potentially revealed clear AGN signatures in the rest-UV in LRDs for the first time, via a detection of a strong Ly$\alpha$ emission and a broad \mgii\, doublet. We present the data in \autoref{sec:data}, line and continuum fitting as well as size measurements in \autoref{sec:data_analysis}, galaxy and black hole properties in \autoref{sec:res}, and the final discussion in \autoref{sec:disc_sum}.
\par
Throughout this work we assume a flat $\Lambda$CDM cosmology \citep[e.g.][]{planck20} with $\Omega_{\mathrm{m},0}=0.3$, $\Omega_{\mathrm{\Lambda},0}=0.7$ and H$_0=70$ km s$^{-1}$ Mpc$^{-1}$, and a \citet{chabrier} initial mass function (IMF) between $0.1-100$ $M_{\odot}$. All magnitudes are expressed in the AB system \citep{oke74}.

\begin{figure*}
\begin{center}
\includegraphics[width=.99\textwidth]{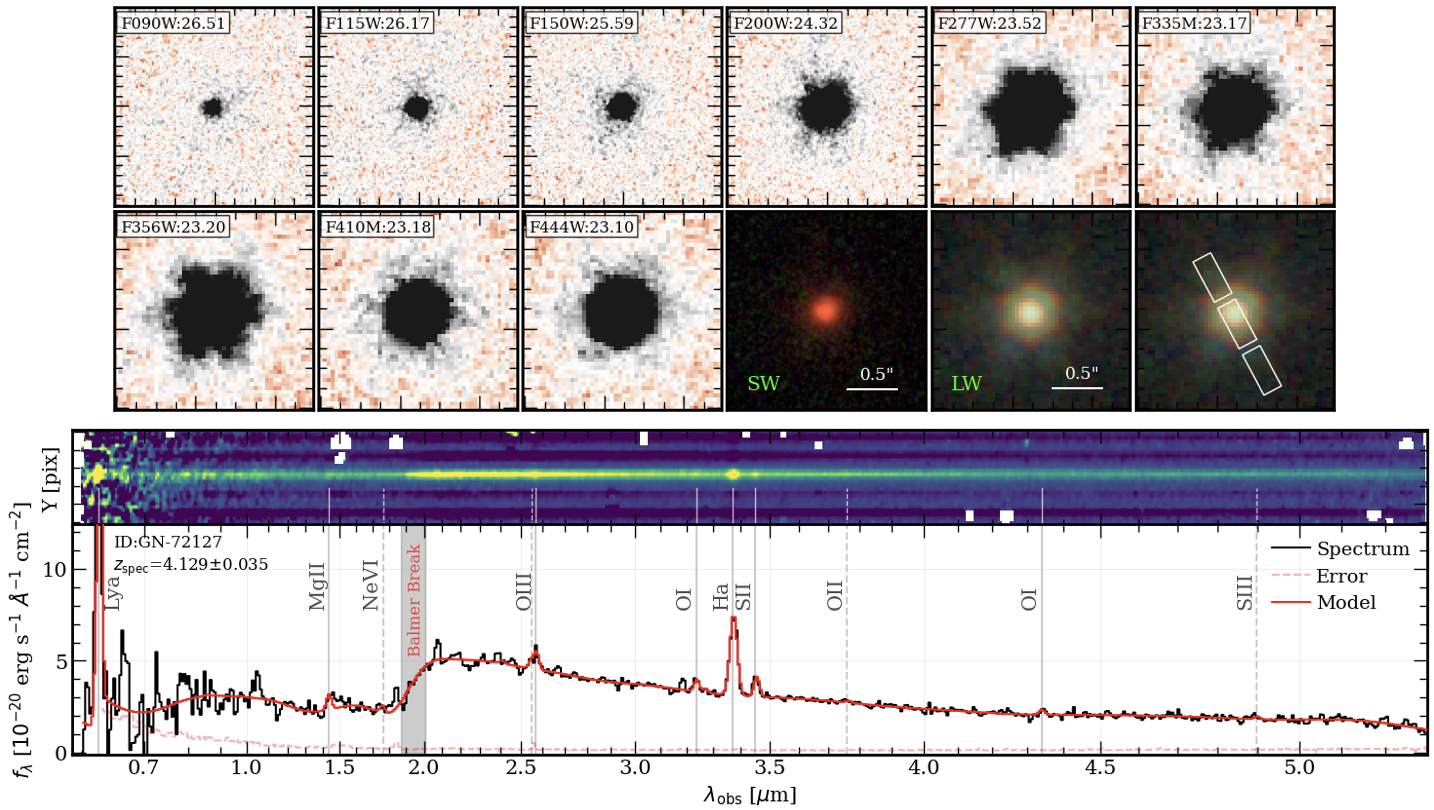}
\caption{\textbf{Top:} \textit{JWST}/NIRCam 2\farcs{0} stamps and the RGB short and long wavelength color images comprised of the F115W, F150W, F200W and F277W, F356W, and F444W bands, respectively. The MSA slitlet layout (white) is overlaid on a separate LW color image, for clarity. The source has a very clear PSF-dominated morphology present in all filters. Each panel shows the total magnitude as presented in the JADES DR3 catalog \citep{deugenio24}. The source is extremely bright and is detected in all bands at $>50\sigma$.
\textbf{Middle:} 2D MSA PRISM spectrum.
\textbf{Bottom:} Optimally extracted 1D spectrum of the galaxy in the observed frame. We show the data in black, while the uncertainty on the spectrum is in dashed red. Best-fit spline continuum and line \textsc{msaexp} model to the data is shown in solid red.
Assuming the best-fit to the PRISM with $z_{\rm spec}$ = $4.129\pm0.035$, we show the positions and label the prominent emission with significant ($>3\sigma$) detections as solid vertical lines. 
Emission lines for which we only obtain an upper limit are shown with dashed lines. The position of the Balmer/4000 \AA\, break is shown with a shaded region.}
\label{fig:fig1}
\end{center}
\end{figure*}

\section{Observations and Data} \label{sec:obs_data}
\label{sec:data}
In our work we use the data obtained as a part of the \textit{JWST} Advanced Deep Extragalactic Survey (JADES; \citealt{bunker23};
\citealt{eisenstein23,eisenstein23b}), which is based on more than 770 hours of the GTO time with the NIRCam and NIRSpec instruments. In particular, we focus our efforts on the public JADES Data Release 3 (DR3), which covers the GOODS-North field \citep{giavalisco04} and includes the NIRCam coverage in 7 broad-band (F090W, F115W, F150W, F200W, F277W, F356W, F444W) and 2 medium-band (F335M, F410M) filters. The data are fully publicly available, including the reduced spectra and catalogs\footnote{\url{https://archive.stsci.edu/hlsp/jades}}.

Most pertinent to our work, this field was targeted by NIRSpec Multi-Shutter Array (MSA) over 4 separate MSA observations. Briefly, each observation consisted of 3 pointings, each with a small spatial offset and used the PRISM ($R\sim100$), G140M, G235M and G395M ($R\sim1000$) gratings. Observations used in our work were completed between
April and May 2023. Details concerning both the photometric and spectroscopic data reduction, calibration and extraction of the spectra can be found in \citet{bunker23} and \citet{deugenio24}.

We, however, would like to note a few crucial details. Factors such as path-loss correction, flux calibration uncertainty, slit position, source location within the slit, and source self-subtraction can affect the normalization, especially when combining observations from multiple MSA pointings. To address this, the JADES team has rescaled the spectra  by convolving the 1D spectra with all NIRCam filters and correcting them by fitting a first-order polynomial to the ratio between NIRCam and convolved NIRSpec flux densities (similar to \citealt{greene24,kokorev23c}; Price et al., in prep). We verified the consistency between photometry \citep{deugenio24} and spectra for all dispersers, and find the results satisfactory.

Second, the MSA configuration employed by the JADES team allows for the contamination of the spectra by other objects in the same detector row. However, for bright objects with multiple line detections, and robust spectroscopic redshifts, the contaminant lines are easy to identify and mask out.

\begin{deluxetable}{cc}[]
\tabcolsep=2mm
\tablecaption{\label{tab:tab1}
	Source Properties}.
\tablehead{}
\startdata
R.A. & $189.265718$ \\
Dec & $62.168393$\\
$z_{\rm spec}$ (PRISM) &  $4.129\pm0.035$\\
$z_{\rm spec}$ (G235M) &  $4.1306\pm0.0004$\\
$z_{\rm spec}$ (G395M) &  $4.1295\pm0.0005$\\
$A_{\rm V}$ (H$\alpha$/H$\beta$) [mag] & $1.1^{+0.8}_{-0.1}$ \\
$r_{\rm eff, optical}$  [pc] & $<300$ \\
$r_{\rm eff, UV}$   [pc] & $490\pm50$ \\
\hline
\multicolumn{2}{c}{FWHM [km/s]} \\
\hline
Narrow Lines (G235M, G395M) & $510\pm72$ \\
Broad Lines (H$\alpha$) & $2573\pm585$ \\
Narrow \mgii\, & $482\pm371$ \\
Broad \mgii\,  & $1100\pm535$ \\
Absorption Lines (Host) & $959\pm162$ \\
\hline
log$_{10}(M_{\rm BH}/M_{\odot})$ (H${\alpha}$) & $7.31\pm0.21$ \\
log$_{10}(M_{\rm BH}/M_{\odot})$ ($L_{5100}$) & $7.90\pm0.15$ \\
log$_{10}(M_{\rm dyn}/M_{\odot})$ & $<10.76$\\
log$_{10}(\Sigma_*/M_\odot$kpc$^{-2}$)  & $<11.01$\\
$L_{\rm bol}$ [erg s$^{-1}$] & $(3.46\pm 1.10)\times10^{44}$ \\
Ly$\alpha$ EW$_{\rm 0}$ [\AA] & $400\pm$50\\
\hline
\multicolumn{2}{c}{BAGPIPES Non-Parametric SFH Fit} \\
\hline
$A_{\rm V}$ (continuum) [mag] & $0.43\pm0.08$ \\
log$_{10}(M_{*}/M_{\odot})$ & $10.63\pm0.05$\\
Mass-weighted Age [Gyr] & $0.74^{+0.21}_{-0.29}$\\
SFR$_{\rm 10}$ [$M_\odot$/yr] & $0.036^{+0.818}_{-0.025}$ \\
SFR$_{\rm 100}$ [$M_\odot$/yr] & $0.070^{+0.847}_{-0.047}$ \\
$Z/Z_{\odot}$ & $0.97\pm0.16$ \\
log$_{10}(U)$ & $-3.07\pm0.13$ \\
\enddata
\end{deluxetable}

\begin{deluxetable}{ccc}[]
\tabcolsep=2mm
\tablecaption{\label{tab:tab2}
Measured emission line fluxes $^\dagger$ and rest-frame equivalent widths of absorption lines (EW$_0$)}.
\tablehead{Line & $\lambda_{\rm rest}$ [\AA] & Flux [10$^{-20}$ erg s$^{-1}$ cm$^{-2}$]}
\startdata
Ly$\alpha$ & 1215.4 & $2875.8\pm288.1$ \\
\mgii\, (narrow) & 2796.5 & $122.2\pm60.0$ \\
\mgii\, (broad) & 2796.5 & $90.1\pm40.2$ \\
\mgii\, (narrow) & 2803.1 & $142.5\pm60.5$ \\
\mgii\, (broad) & 2803.1 & $225.8\pm95.2$ \\
$[$\neiii\,$]$ & 3867.9 & $123.5\pm40.5$ \\
$[$\oiii$]$ & 4363.0 & $108.4\pm25.5$ \\
H$\beta$ (narrow) & 4862.7 & $105.5\pm55.0$\\
$[$\oiii\,$]$ & 4959.5 & $155.2\pm32.2$\\
$[$\oiii\,$]$ & 5007.2 & $307.7\pm35.5$\\
OI & 6302.2 & $169.8\pm30.5$ \\
\nii\, & 6549.0 & $293.0\pm25.6$ \\
H$\alpha$ (narrow) & 6562.8 & $444.0\pm94.1$\\
H$\alpha$ (broad) & 6562.8 & $742.8\pm150.0$\\
\nii\, & 6585.2 & $625.7\pm77.4$ \\
He\,\textsc{i} & 6680.0 & $72.7\pm23.2$ \\
\sii\, & 6716.5 & $104.9\pm30.0$ \\
\sii\, & 6731.8 & $80.2\pm25.8$ \\
O\,\textsc{i} & 8446.0 & $53.0\pm17.5$ \\
\hline
\multicolumn{3}{c}{Absorption Lines} \\
\hline \hline
Line & $\lambda_{\rm rest}$ [\AA] & EW$_0$ [\AA] \\
\hline
H$\eta$ & 3835.4 & $7.0\pm0.6$ \\
H$\zeta$ & 3889.1 & $6.6\pm1.0$ \\
Ca K & 3934.0 & $1.4\pm1.1$ \\
H$\epsilon$ & 3970.1 & $12.0\pm1.6$ \\
H$\delta$ & 4101.7 & $4.1\pm1.9$ \\
H$\gamma$  & 4340.4 & $5.5\pm1.5$ \\
H$\beta$ (absorption) & 4862.7 & $16.3\pm4.4$\\
H$\alpha$$^1$ (absorption) & 4862.7 & $1.41\pm0.5$\\
\enddata
\begin{tablenotes}
\footnotesize{$^\dagger$ No dust correction was applied.} \\
\footnotesize{$^1$ Derived from the best-fit \textsc{BAGPIPES} stellar model.} \\
\end{tablenotes}
\end{deluxetable}

\section{Data Analysis} \label{sec:data_analysis}
The data from JADES in the GOODS-N field are excellently suited to explore the enigmatic rest-frame UV emission of compact obscured AGN. The available \textit{JWST} photometry covers a complete wavelength range from 0.4 -- 5 $\mu$m, in 7 broad and 2 medium bands, reaching a median 5$\sigma$ depth of 29.4 AB mag in F444W filter allowing for robust target pre-selection. Moreover, the exquisite coverage provided by all 3 medium resolution dispersers, especially by the bluest G140M grating, and the PRISM allows for robust identification of spectral features such as broad and narrow lines in both rest-UV and rest-optical parts of the spectrum.

\subsection{Target Selection}
\label{sec:targsel}
In the last year many BL AGN have been successfully identified via their broad band colors and rest-frame optical compactness, and subsequently verified via the presence of broad lines in their rest-optical spectra \citep{greene24, furtak24, killi23, kocevski23, kokorev23c, wang24}.
Namely, using a combination of broad-band NIRCam colors covering the characteristic "break" feature present in LRDs and 
a compactness criterion was shown to be extremely successful ($>80 \%$) at identifying broad line AGN with photometry and size alone.

To identify our targets we use the same criteria as \citet{labbe23}, \citet{greene24}, and \citet{kokorev24}, aiming to maintain a consistent approach in selecting LRDs, particularly regarding aperture sizes.
These are as follows: 
\begin{equation*}
\begin{aligned}
\texttt{red} \, \texttt{1} & = \mathrm{F115W}-\mathrm{F150W}<0.8 \quad \& \\
&\mathrm{F200W}-\mathrm{F277W}>0.7 \quad \& \\
&\mathrm{F200W}-\mathrm{F356W}>1.0  
\end{aligned}
\end{equation*}
or
\begin{equation*}
\begin{aligned}
\texttt{red} \, \texttt{2} & = \mathrm{F150W}-\mathrm{F200W}<0.8 \quad \& \\
&\mathrm{F277W}-\mathrm{F356W}>0.6 \quad \& \\
&\mathrm{F277W}-\mathrm{F444W}>0.7,
\end{aligned}
\end{equation*}
plus the compactness cut given by:
\begin{equation*}
\texttt{compact} = f_{\rm f444w} (0\farcs{4})/f_{\rm f444w} (0\farcs{2})<1.7.
\end{equation*}
The final selection is then ($\texttt{red} \, \texttt{1} \, | \,\texttt{red} \, \texttt{2}$) \& \texttt{compact}.

The JADES GOODS-N DR3 public photometric catalog includes both circular aperture and total (i.e., Kron) photometry for 85,709 sources in the GOODS-N field \citep{deugenio24}. Since we are focusing on compact sources, we specifically use $D=$0\farcs{3} apertures (\texttt{CIRC2}) and convert them to total fluxes using the aperture corrections provided by the JADES team. These smaller apertures offer a good balance for reliably measuring fluxes of point-like objects \citep[e.g., see][]{kokorev23c, kocevski24, labbe23, trussler24}, while keeping aperture corrections relatively low. Finally, we also require for the object to be covered by all 3 medium resolution gratings and the PRISM, and also have a valid redshift from the JADES catalog \citep{deugenio24}.

Using this selection method, we first identify 11 LRDs in the data ranging from $z\sim4-7$, all targets have clear rest-optical line detections in the PRISM, as well as the G235M and G395M gratings. Moreover, 4/11 LRDs that we find were already confirmed and presented as BL AGN in a work by \citet{maiolino23b}. Not all however have significant line detections in the rest-UV, covered by the G140M grating and PRISM. This is imperative to our study as we aim to simultaneously examine the rest-UV and rest-optical properties of LRDs. Despite that, we find that a single source from the initial 11 stands out in particular, owed to its broad lines in the rest-optical and multiple line detections in the rest-UV.

We thus identify \srcname, a potential AGN candidate at $z=\zsource$, located at R.A. = 189$^{\circ}$.265718, decl.= 62$^{\circ}$.168393 (\autoref{tab:tab1}). Consistent with LRDs selected in previous studies, \srcname\, is red in the rest-optical but remains visible, albeit moderately red, in the rest-UV. It is also very bright in the F444W broad-band filter ($\sim$ 23.1 mag) while remaining completely point-like (see \autoref{sec:size}). 
Of prime interest to our work, however, is the spectral coverage of \srcname. A combination of NIRSpec/PRISM observations and, most crucially, the exquisite coverage by three medium-resolution dispersers ($t_{\rm int}$=3107 s each)
provides the most comprehensive set of data to date for investigating the physics of these reddened AGN. We present 2\farcs{0} cutouts of the source in each medium and broad-band NIRCam filter alongside a PRISM spectrum in \autoref{fig:fig1}.

Despite clear AGN signatures in LRDs, a vast majority of them remains X-ray undetected \citep[e.g. see ][]{akins24,greene24,maiolino24,pacucci24b}. In line with this, we also find nothing at the position of the source in the deep ($\sim170.43$ ks) Chandra data \citep{chandra_gdn}.

\subsection{Initial Line Identification}
We start our investigation of all available low and medium resolution spectra by running a heavily modified version of \textsc{msaexp} \citep{msaexp}. Briefly, \textsc{msaexp} models emission and absorption lines as Gaussians and fits the continuum as a series of cubic splines, albeit at a fixed (and user defined) velocity width for all lines.
Our key modifications\footnote{A modified code is available upon request.} to the code include making the line width a free parameter, and also adding an option to fit the same line with multiple components (i.e., narrow and broad). For each available spectrum of the source, we allow the fit to search for the best fit within a narrow redshift range ($\Delta z\sim0.02$), with the prior set by drawing from the JADES spectroscopic catalog. We overlay the best fit to the PRISM in \autoref{fig:fig1}.

We find that the $z_{\rm spec}$ derived from the PRISM and 3 medium resolution dispersers match up quite well, without any significant offsets ($<150$ km/s). With this approach we identify a number of the key features. 
These are - a potential broad \mgii\, and H$\alpha$ emission, strong [\oiii] and [\nii] doublets, a strong Balmer break as well as the higher order (H$\gamma$ through H$\eta$) Balmer absorption features (\autoref{fig:fig2}). The latter two imply that the rest-optical continuum could be dominated by an older stellar population. All of the above are securely identified in PRISM as well as the gratings. We also note a very prominent Ly$\alpha$ emission, albeit only in the PRISM, which has insufficient spectral resolution to compute any systemic velocity offset.

We note that not all spectral lines can be modeled well by our routine. While it is adequate at fitting the majority of narrow and broad spectral features rather quickly, which is useful for large samples of spectra, its reliability is quickly diminished by intricacies of certain line complexes. This is especially apparent when uncommon line ratios are required, in the regions where S/N is low or the continuum is simply too noisy.
A robust identification of certain spectral features, such as broad lines, is imperative to properly classify the object. Therefore, we will focus on certain parts of the spectrum with a more sophisticated procedure, which we will describe in subsequent sections.

\subsection{G395M: Broad H$\alpha$ +[\nii]}
\label{sec:g395m}
Compared to the PRISM the G395M grating has a much higher spectral resolution which allows us to resolve the, otherwise blended, emission from the [\nii] doublet and H$\alpha$. The [\sii] doublet remains blended however. Crucially, the availability of multiple spectral pixels sampling the line complex enables us to try and fit both broad and narrow components to the Balmer emission. While we do not detect any contaminant lines in the spectrum we note the presence of corrupted data (NaN values) in the spectrum, likely a result of issues during data reduction or 1D spectrum extraction. These features are, however, minor and do not affect our fitting routine. In addition, as we already mentioned in \autoref{sec:data}, due to the MSA configuration employed by the JADES team, the spectra can overlap, resulting in additional lines from neighboring sources. However, as the redshift for the source is known precisely, these contaminant lines are easily masked out.

We fit each line complex simultaneously alongside the continuum. Each emission line is modeled with a standard Gaussian profile, with the position allowed to vary within a narrow redshift range around the best-fit \textsc{msaexp} solution. We fix the ratio between the [\nii$_{\lambda6549}$] and [\nii$_{\lambda6585}$] to 1:3, respectively, as is normally done in the literature \citep[e.g. see][]{de_graaff24}. We also fit the H$\alpha$ emission with both narrow and broad components. The ratio between lines in the [\sii] doublet is allowed to vary freely. All narrow line velocities are tied together and we assume the same redshift for all lines. Finally, we model the local continuum with a first order polynomial. Given the limited number of spectral pixels we do not include an absorption component when we fit the H$\alpha$, as the fit would become too degenerate. Instead, to account for H$\alpha$ absorption, we use the best-fit stellar continuum model fitted to the host galaxy as described in \autoref{sec:host}. The derived line fluxes of H$\alpha$ and [\nii] lines are then corrected for the stellar absorption. Generally however, the EW of the stellar absorption is comparable for all Balmer lines, meaning that the redder (i.e. H$\alpha$) absorption lines have a weaker contribution to the spectra, compared to the higher order lines, and are thus less important to take into account.

The fit is initialized by creating all of the models on an oversampled grid, which is then interpolated onto the wavelength axis that mimics the heterogeneous grating resolution. Finally, we allow for a custom over sampled spectral resolution scaled by a factor of 1.3. It was shown that a spectral resolution of a point-like source can be higher than that of a source uniformly illuminating the slitlet \citep[see e.g.][]{de_graaff23,greene24,kokorev23c}. The best fit is a found via a nonlinear least-squares minimization, with the MCMC uncertainty derived from the covariance matrix.

From the fit we derive a redshift of $z_{\rm spec}=4.1294\pm0.004$, FWHM of the narrow lines (H$\alpha$, [\nii], [\sii])  equal to $510\pm72$ km/s and a broad H$\alpha$ component with a width of $2573\pm585$ km/s. We strongly detect the [\nii] and [\sii] doublets, and recover both narrow and broad $H\alpha$ components at $~3\sigma$ level. To verify whether a broad line fit is required at all, we perform additional modeling without allowing for a broad H$\alpha$ emission. To verify the significance of this broad-line we calculate the Bayesian Information Criterion (BIC) difference between the narrow+broad and narrow lines-only fits. We find the narrow only fit to be an inadequate representation of the data (BIC$_{\rm narrow}-$BIC$_{\rm narrow+broad}$=$\Delta$BIC$>20$), which indicates a very strong evidence \citep[using the criteria defined in][]{jeffreys61} for the broad component being present in H$\alpha$.

\begin{figure*}
\begin{center}
\includegraphics[width=.99\textwidth]{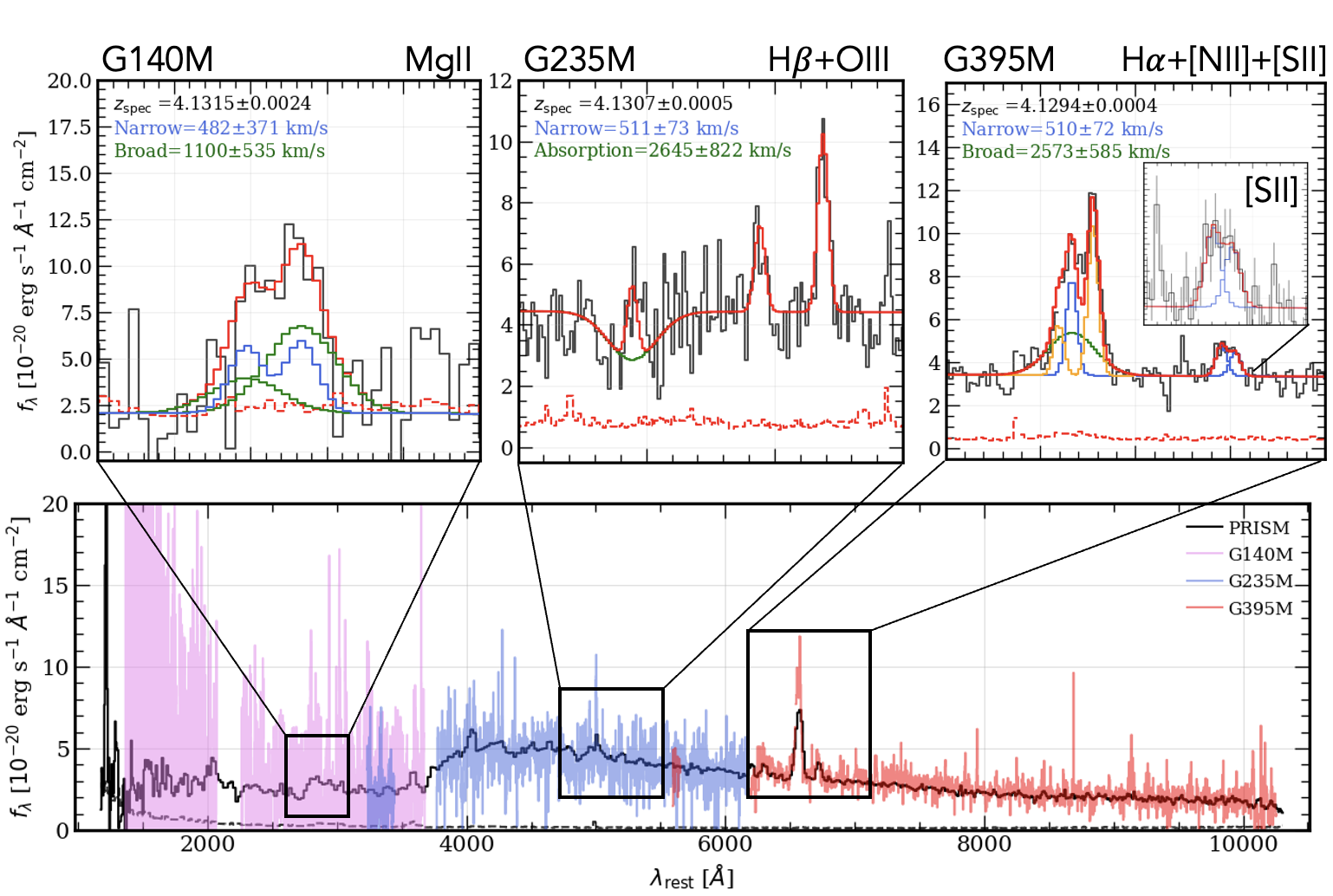}
\caption{\textbf{Top:} A series of panels showing fits to the main line series of interest. From left to right these are,
a tentative broad line fit to the \mgii\, doublet; H$\beta$ and [\oiii] doublet; H$\alpha$, [\nii] and [\sii] doublets. We show best-fit narrow lines in blue, apart from the [\nii] which is shown in orange for clarity. The broad emission (or absorption) best fits are shown in green. The total combined model including the continuum is plotted in red. The original spectrum is in black and the error spectrum is dashed red. 
\textbf{Bottom:} Full collection of all available medium and low resolution MSA spectra. These are as follows, PRISM (black), G140M (violet), G235M (blue) and G395M (red). The presence of broad lines in the G140M and G395M spectra is our first hint of AGN activity.}
\label{fig:fig2}
\end{center}
\end{figure*}

We measure log$_{10}$([\nii]$_{\lambda6586}$/H${\rm \alpha}$)=$0.15^{+0.24}_{-0.16}$, which is too high to have originated from star-formation \citep{kewley01,grossi09}. The [\sii] doublet at $\lambda\lambda$6717,6731 \AA\, is another useful diagnostic for gas ionization when coupled with the adjacent H$\alpha$ line. We also find it to be elevated, with log$_{10}$([\sii]$_{\lambda\lambda6717,6731}$]/H${\rm \alpha}$)=$-0.38^{+0.21}_{-0.13}$. While it is possible to reach these high line ratios via e.g. shocks, the presence of a robust broad H${\alpha}$ emission exceeding 2000 km/s, much faster than could be explained by shocks \citep{allen08}, gives us the first hint as to the presence of an AGN \citep{kewley06} in this otherwise seemingly stellar-light dominated galaxy. The best-fit is presented in the rightmost panel of \autoref{fig:fig2}.
For now we will refrain from making a definitive statement regarding the AGN presence and will explore other gratings.

\subsection{G235M: H$\beta$ and [\oiii]}
\label{sec:g235m}
We continue our investigation by shifting to a bluer medium resolution disperser - G235M (middle panel of \autoref{fig:fig2}). Following the same fitting procedure as before, we now fit the H$\beta$ line with two components (emission and absorption), and the [\oiii] where we do not fix the ratio between the lines. We assume the same velocity dispersion for all emission lines and leave the velocity of the absorption line to be independent. The H$\beta$ is fit with only a single narrow component. 

Firstly, we derive a $z_{\rm spec}=4.1307\pm0.005$, consistent with the G395M, FWHM of the emission lines equal to $529\pm81$ km/s and a largely unconstrained broad absorption component with a width of $2429\pm835$ km/s. We strongly detect the narrow [\oiii] doublet and find the FWHM to match that of the narrow lines in the G395M grating. Therefore it is likely that H$\alpha$, H$\beta$, [\nii], [\oiii] and [\sii] originate from the same region in the galaxy. We also confirm a very weak H$\beta$ emission, marginally detected at $\sim1.5\sigma$ level. At the same time we also can clearly detect an H$\beta$ line in absorption, however the noise present around that feature does not allow us to securely identify the width of that feature. 
While uncertain, what is clear is that the absorption profile is wider compared to the emission lines. We will focus on all available absorption features and their widths in the next sections.  Notably, while the narrow H$\beta$ is largely undetected we can still use this information and derive the lower limit on the  log$_{10}$([\oiii]$_{\lambda5007}$/H${\rm \beta})>0.50$. This high value again puts us at odds with having originated purely from star formation.

\subsection{G140M: \mgii\,}
Finally we focus on the last available grating the G140M. As is typical of LRDs, the object is rather faint in the rest-UV as such we do not detect a large array of lines in this disperser compared to the other ones, however one feature stands out. We perform a fit to the \mgii\, doublet ($\lambda\lambda 2796,2803$) with a narrow and broad component for each, as shown in the left panel of \autoref{fig:fig2}. During the fit we allow the \mgii\,$_{2796}$/\mgii\,$_{2803}$ ratio to vary freely. We find the FWHM of the narrow component to be equal to $482\pm371$ km/s, not constrained as well as for the other narrow lines in the spectrum, but still consistent within the uncertainty. Including the broad component, we find its width to be equal to $1100\pm535$ km/s, marginally detected at $\sim2\sigma$ level. As before we attempt the same fit with a narrow only \mgii\, doublet, and find it to be moderately worse ($\Delta$BIC=2.5), compared to the narrow+broad fit. We also find that the \mgii\,$_{2796}$/\mgii\,$_{2803}$ for both narrow and broad lines is consistent with the accepted ranges $\sim 0.5 - 1$ within the uncertainty \citep{laor97,wang09}.

While the evidence is not as strong as for the broad H$\alpha$, we will refrain from making a definitive statement on whether the broad \mgii\, emission is present in the rest-UV spectrum of the source. This possibility however can not be ruled out either, as it does appear that a broad-line component makes the fit better. The potential presence of broad \mgii\, emission provides us with another piece of the evidence that an active black hole might be present in \srcname. 

Just like the broad hydrogen features, broad \mgii\, lines are also formed in the BLR, but typically originate slightly farther out from the central black hole compared to e.g. H$\alpha$. This difference in location can result in somewhat lower velocities for \mgii-emitting gas clouds \citep[see e.g.][]{wang09}, which is what we find in our fit. In addition, the S/N of the G140M at the position of \mgii\, barely allows us to model the continuum, as such broader \mgii\, wings are obscured by the noise, making our derived line width lower than what it is in reality.

\begin{figure}
\begin{center}
\includegraphics[width=.45\textwidth]{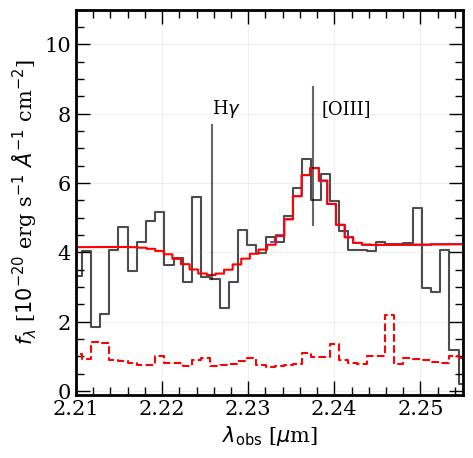}
\caption{A fit to the H$\gamma$ and [O\,\textsc{iii}]$\lambda4363$ lines. The spectrum is in black and the total fit is in red. The error spectrum is shown in dashed red. The velocity widths and redshift were fixed to the values form the previous fits. The derived intrinsic [\oiii]$\lambda4363$/[\oiii]$\lambda5007$ of $\sim0.5$ is indicative of extremely high ionization, likely induced by AGN activity. }
\label{fig:fig3}
\end{center}
\end{figure}

\subsection{Balmer Decrement}
\label{sec:decr}
It is well established that the ratio between observed fluxes of Balmer series lines can
be used to determine the dust extinction. Given the quiescent nature of the source we have a strong reason to believe that all observed emission lines originate from either the NLR or the BLR around an AGN. As such applying the dust attenuation derived from the continuum would not be adequate to derive AGN properties. To compute the $A_{\rm V}$ we use the narrow line ratio - H$\alpha$/H$\beta$ computed from  our fits to the G395M and G235M dispersers. As mentioned in \autoref{sec:g235m}, the narrow H$\beta$ line is only detected at $2\sigma$, so our $A_{\rm V}$ would only be an approximation.  By doing this we find H$\alpha$/H$\beta$ $\approx 4.28$. Provided the same ratio holds between narrow and broad emission lines, we also confirm that the broad component in H$\beta$ would not be detectable in our data (EW$_0\sim5$ \AA), thus justifying our narrow only fit to H$\beta$ in \autoref{sec:g235m}.

In line with other works on reddened quasars \citep[e.g. see][]{hopkins04}, high-$z$ galaxies \citep{capak15,reddy15,reddy18}, and LRDs themselves \citep[e.g.][]{furtak24,greene24,kokorev23c} we use the Small Magelannic Cloud (SMC) reddening law \citep{gordon03}. Assuming that case B recombination applies in the NLR of an AGN \citep{osterbrock89}, we find that the observed line ratio gives us a moderate attenuation of the region around the central black hole equal to $A_{\rm V}$ $=1.1^{+0.8}_{-0.1}$.

\begin{figure*}
\begin{center}
\includegraphics[width=.99\textwidth]{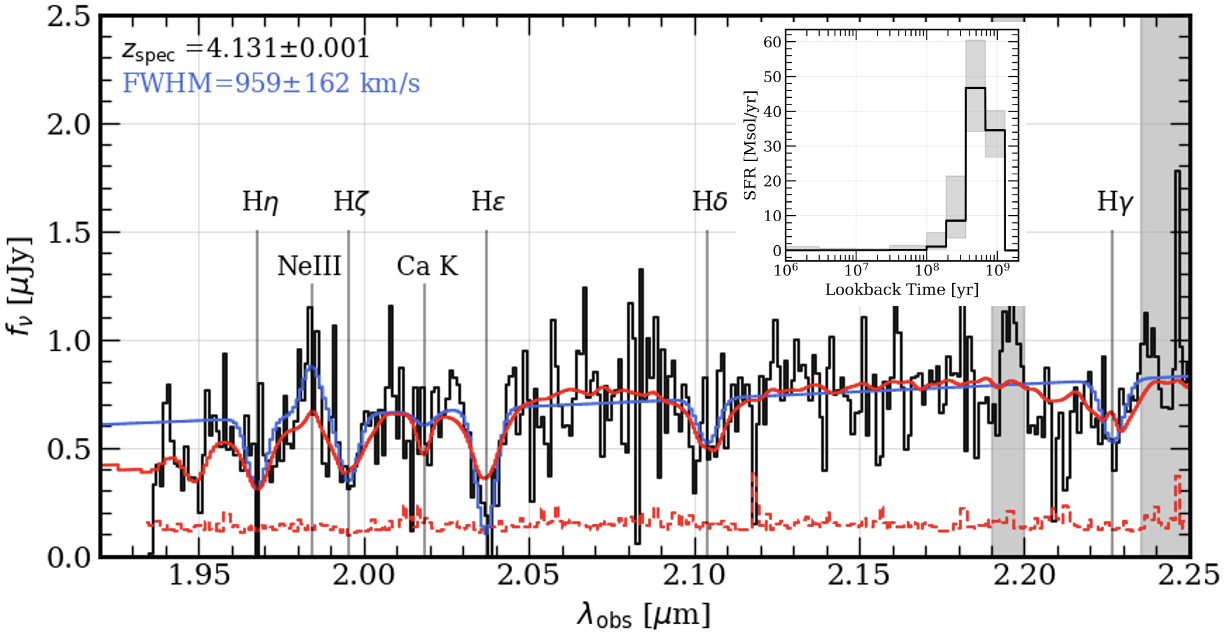}
\caption{Examination of the absorption features in the G235M spectrum of the quiescent host galaxy. We show our best fit to the absorption lines+[\neiii] in blue, while the non-parametric SFH BAGPIPES fit is in red. The regions of the spectrum (black) that were masked out due to contaminant lines are shown as a shaded area. The error array is in dashed red. This galaxy exhibits deep Balmer absorption lines with the spectrum resembling that of an A-type star, common in lower-redshift post-starburst galaxies \citep{goto07,wild09,wild20}, as well as most recent exploration of $z>4$ quiescent galaxies \citep{carnall23,de_graaff24}. On the inset we show the best-fit non-parametric SFH, which clearly shows that \srcname\, experienced a significant, rapid drop in the SFR within the last $\sim200-300$ Myr ($z\sim5.8$).}
\label{fig:fig4}
\end{center}
\end{figure*}

\subsection{Other Notable Lines}
Beyond the lines discussed in earlier sections, the spectra reveal a multitude of other emission and absorption features. As these are quite numerous, considering the volume of data for this object alone, we will not focus on each one individually, but rather point out some that could be of particular interest to our study. The derived fluxes for all lines in our work can be found in \autoref{tab:tab2}.

We fit the Ly$\alpha$ emission in the PRISM spectrum using a single Gaussian model, allowing both the FWHM and redshift to vary. Due to a limited spectral resolution of the PRISM at this wavelength ($R\sim60$), we are unable to precisely measure the width or offset of this line from the systematic redshift. However, the line spans 3 spectral pixels, each approximately 900 km/s wide in the rest frame, suggesting that the Ly$\alpha$ emission in \srcname\ could be broad. Notably, the line is quite strong, with a measured rest-frame equivalent width (EW$_{0}$) of $400\pm50$ \AA, significantly exceeding the typical values found in star-forming galaxies at $z\sim4.5$ \citep{zheng14,hashimoto17}.

In the G235M spectrum we identify the auroral [\oiii] $\lambda4363$ line, which can be used to trace both ionization and metallicity. It has been suggested that at higher metallicities [\oiii] can blended with the adjacent [Fe\,\textsc{iii}]$\lambda4360$ line \citep[e.g. see][]{curti17}. We attempt two fits, one including the [Fe\,\textsc{iii}] line and one without it. In both cases we fix the velocity widths and redshift to the other narrow lines. Our results indicate that the [\oiii] line by itself better represents the data, as opposed to fitting it alongside [Fe\,\textsc{iii}] ($\Delta$BIC$\sim3$). Given that there are only so many spectral pixels covering the line, we believe that the fit with a $\lambda4360$ feature line is simply too degenerate to yield an adequate BIC. Therefore, we cannot entirely rule out the possibility that [\oiii]$\lambda$4363 is free from contamination. This is well reflected in our uncertainties on the derived line flux. We show the best fit to the line in \autoref{fig:fig3}.

Finally, we note the presence of the O\,\textsc{i} $\lambda8446$ line detected in the PRISM spectrum at $\sim3\sigma$ significance \autoref{tab:tab2}. The O\,\textsc{i} line is produced through the Ly$\beta$ fluorescence process which involves the absorption of Ly$\beta$ photons by neutral oxygen, which is then re-emitted at 8466 \AA. This process requires dense, highly ionized environments, such as the BLR of an AGN \citep{osterbrock06}.

\subsection{Exploring the Host Galaxy}
\label{sec:host}
So far we have identified broad line emission, enhanced emission line ratios and a high Ly$\alpha$ EW which all point toward a presence of an AGN in \srcname. Despite that, the unambiguous presence of the Balmer break and associated absorption lines in the spectrum clearly indicates that the majority of the continuum emission in \srcname\, is still likely dominated by older stars. To this end we would like to perform a dedicated fit to the host galaxy in order to derive the properties of the stellar population in \srcname.

The available \textit{JWST} PRISM and medium grating data are jointly fit with the BAGPIPES SED fitting code \citep{carnall18,carnall19}. The setup for the fit is as follows. We adopt a non-parametric SFH with the `bursty continuity' prior from \citet{tacchella22}, using eight time bins where the SFR is fit to a constant value in each bin. The first four bins are set to lookback times of 0-3 Myr, 3-10 Myr, 10-30 Myr, and 30-100 Myr, while the last four bins are logarithmically spaced between 100 Myr and $t_\mathrm{max} = t_\mathrm{universe}(z=z_\mathrm{spec}) - t_\mathrm{universe}(z=20)$. Dust attenuation is assumed to follow the \citet{calzetti00} law and the rest-frame V-band attenuation is fit (log-uniform prior) in the range $A_{\rm V} = 0.001 - 5$ mag. Stellar nebular emission is included with an ionization parameter in the range $-4 \leq$ log $U \leq -2$. We allow for a stellar velocity dispersion in the range 50--500 km/s and metallicity between 0.1--2.5 $Z_\odot$ (log-uniform prior for both). 
Finally, we allow BAGPIPES to modify the input spectral noise array by a factor between 0.5--2. As mentioned above, the AGN-like ratios of the narrow emission lines imply that they potentially originate in the NLR, rather than the host. We therefore have removed the contribution of the emission lines to our fit by masking them out.

From the BAGPIPES fit we find the galaxy to be very massive log$_{10}(M_*)=10.63\pm0.02$, with little dust attenuation ($A_{\rm V}\sim0.4$) and essentially devoid of all star-formation activity (log$_{10}$(sSFR)$<-11.5$). 
The best-fit SFH reveals a short burst of star-formation some $\sim 300$ Myr ago, which lasted for $\sim 100$ Myr, forming at 
roughly $40-60$ $M_\odot$/yr. The galaxy then appears to have quenched rapidly at $z\sim5.5$, remaining largely dormant all the way until the time we observed it. In \autoref{fig:fig4} we present the best-fit BAGPIPES SED, alongside a line-specific fit to the relevant part of the spectrum containing absorption lines. An array of high order Balmer absorption lines from H$\eta$ to H$\gamma$, alongside a tentative detection of the Ca K feature is reminiscent of the highest-$z$ quiescent galaxies found with \textit{JWST} \citep{carnall23,de_graaff24}.

While \textsc{BAGPIPES} measures the width of the absorption features alongside other parameters, that fit was done on data that span multiple spectral resolutions, thus the derived widths might be inadvertently broadened.
To keep all of our velocity measurements consistent, we have re-fit all of 
the absorption features, as well as the \neiii\, emission with the same method as in \autoref{sec:g395m}. We show our final result in \autoref{fig:fig4}. For the absorption lines we derive a FWHM
of $959\pm162$ km/s. This corresponds to the velocity of the stars in the galaxy, rather than the gas, as was done in previous works concerning BL AGN \citep{maiolino23b}. The absorption lines from stars trace the dynamical mass, and are broader than the narrow lines (FWHM/2.355 = $\sigma_*=407\pm71$ km/s), consistent with the notion that all our narrow emission lines likely originate from the NLR, rather than the host galaxy. 

\begin{figure*}
\begin{center}
\includegraphics[width=.99\textwidth]{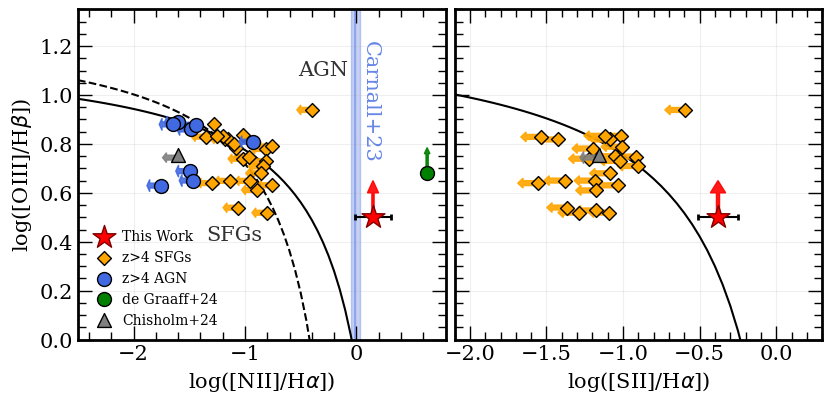}
\caption{BPT (\textbf{left}) and BPT-[\sii] (\textbf{right}) ionization diagnostics. The [\nii] and [\sii] to H$\alpha$ line ratios found in the object (red) appear to vastly exceed those found in high-$z$ SFGs \citep[orange;][]{cameron23,sanders23}, Type I \citep[blue;][]{maiolino23b} and Type II AGN \citep[grey;][]{chisholm24}.
On the other hand we find a comparable ionization to the highest-$z$ QG from \citet{de_graaff24} (green) and \citet{carnall23} (vertical blue line). The line ratios we find are
indicative of AGN-dominated ionization as the tracks from \citet{kewley01} (solid) and \citet{kauffmann03} (dashed) appear to suggest.}
\label{fig:fig5}
\end{center}
\end{figure*}

\subsection{Size Measurements}
\label{sec:size}
As the cutouts in \autoref{fig:fig1} indicate, the source is very compact, bordering on unresolved in the redder bands. We now would like to measure the effective radius ($r_{\rm eff}$) to confirm whether that is indeed the case by modeling the object with \textsc{GALFIT} \citep{peng02,peng10}. When fitting we take into account the effects of the PSF, which we have measured empirically from the bright stars in the field. We model the object with a S\'ersic \citep{sersic} profile where the source position, brightness, effective radius, S\'ersic index, and axis ratio as allowed to vary. 

In all LW bands (F277W, F356W, F410M, F444W) the measured effective radius varies from $0\farcs{034}-0\farcs{04}$. In this case a source can be considered to be point-like and unresolved since its effective radius is smaller than the empirical PSF HWHM ($\sim 0\farcs{07}$ for F444W). It is however possible to measure sizes of objects that are smaller that the PSF as shown was shown by other works on compact objects \citep{vanderwel14,labbe23}.
Taking the redshift into account, we derive an upper limit on the physical effective radius to be $\lesssim300$ pc. The point-like nature and size of this source in rest-optical is consistent with a highly concentrated, PSF dominated
objects explored in the recent exploration of high-$z$ reddened AGN \citep{furtak24,kocevski24,kokorev23c,kokorev24,labbe23} as well as the similarly compact highest-$z$ quiescent galaxies \citep{carnall23,de_graaff24}.

On the other hand, we marginally resolve the source in the rest-UV, sampled by all other bluer filters, with the median size across all SW filters equal to $0\farcs{07}\pm0\farcs{01}$, roughly $\times2.5$ larger than the PSF (HWHM$\sim$0\farcs{025}), resulting in $r_{\rm eff}=490\pm50$ pc. We also derive a S\'ersic index $n\approx1.15$, implying a lower degree of central concentration in the rest-UV. As \srcname\, is not resolved in rest-optical it is difficult to say with a high degree of certainty if \srcname\, has a more compact morphology in the redder filters. Although it is entirely feasible that the extended emission present in the rest-UV might imply that this part of the spectrum is dominated by stars, rather than the AGN light. A similar finding has already been reported in a stellar-light dominated LRD with extremely strong Balmer emission (Labb\'e et al., in prep.).

Next, we would like to use the effective radius to gauge the dynamical mass as well as the stellar surface density ($\Sigma_*$). As the stellar light will be dominant at longer, rest-optical wavelengths, this would make the F444W band the most physically constraining for this type of study, as was done in \citet{labbe23}. The F444W band should also trace the region with the lowest amount of dust obscuration.

Using the derived velocity widths of the absorption lines and the F444W size of the object we derive the dynamical stellar mass by following $M_{\rm dyn} = 5 \, r_{\rm eff}\times \sigma_*^2/G$, where $G$ is the gravitational constant, \citep{cappellari06,de_graaff24}. The $\sigma$ we derive in \autoref{sec:host} is explicitly that of the stellar population itself, and should better trace the true $M_*$ (plus gas and dark matter) as opposed to the widths of narrow nebular lines. Given the $r_{\rm eff}<300$ pc and the stellar velocity dispersion -- $\sigma_*=407\pm71$ km~s$^{-1}$ -- we obtain an upper limit on $M_{\rm dyn}<10.76$ $M_\odot$, fully consistent with the $M_*$ we measure from SED fitting. Finally, we derive an upper limit on the stellar mass surface of log$_{10}(\Sigma_*/M_\odot$kpc$^{-2}$)$<11.01$. This is high but does not exceed the maximum limits attainable during an intense burst of star formation \citep{hopkins10,grudic19}, and is consistent with some of the densest recently discovered quiescent galaxies at high-$z$ \citep{carnall23,de_graaff24}.

\begin{figure}
\begin{center}
\includegraphics[width=.48\textwidth]{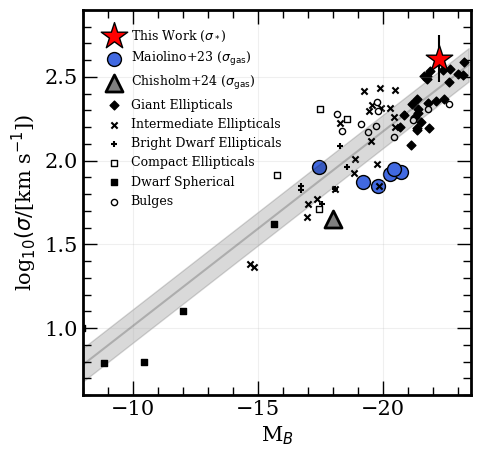}
\caption{An exploration of $z\sim4-7$ AGN on the fundamental Faber-Jackson relation \citep[solid line;][]{faberjackson,shen08,cappellari16}.
We show \srcname\, in red, Type I AGN from \citet{maiolino23b} in blue, and Type II AGN from \citep{chisholm24} in grey. For comparison we additionally overlay results for various local elliptical, spherical and bulge galaxies \citep{bender92}. The source is located exactly on the Faber-Jackson relation, signifying how it is already similar to local ellipticals at $z\sim4$, and further implying that the continuum is indeed stellar dominated.}
\label{fig:fig6}
\end{center}
\end{figure}

\section{Results} \label{sec:res}
\subsection{Black Hole Properties}
Time domain observations of local quasars draw a correlation between the width of the broad Balmer series lines and the size of the BLR \citep[e.g.,][]{kaspi00,greene05}, allowing for the BH mass to be estimated from the width of the emission lines and their luminosities. We compute the BH mass ($M_{\rm BH}$) from the luminosity and the width of the broad H$\alpha$ line, and rest-frame $5100$ \AA\, luminosity of the underlying continuum. Since we have established that the Balmer emission most likely all comes from the region around the AGN itself, i.e. the narrow and broad line regions (NLR and BLR), we will use the $A_{\rm V}$ value derived from the Balmer decrement, and assume that NLR and BLR have similar amounts of dust extinction, to correct the H$\alpha$. On the other hand, the rest-optical continuum is likely to be stellar light dominated, so we will use an $A_{\rm V}\sim0.4$ value derived from the \textsc{BAGPIPES} fit (rather than $A_{\rm V}\sim1.1$ from the decrement) to the host galaxy when correcting the $L_{5100}$.

With the above considerations in mind, from the H$\alpha$ we derive
log$_{10}(M_{\rm BH,H\alpha}/M_{\odot})$=7.31$\pm$0.21 and from the continuum we get log$_{10}(M_{\rm BH, 5100}/M_{\odot})$=7.90$\pm$0.15.
The uncertainty on both of these estimates is largely dominated by the scatter in the relations used to derive the masses, rather than the errors on the line flux and/or dust extinction \citep{kollmeier06}. If the rest-optical was mostly dominated by the AGN emission, one would expect both of these results to be consistent \citep[e.g. see][]{kokorev23c}, however we find that the 5100 \AA\, estimate, even with a comparably modest dust obscuration, returns a mass that is $\times5$ higher. This in return might imply that despite the H$\alpha$ line being clearly detectable, the continuum around it is still dominated by the stellar light. The degree to which this is happening is highly uncertain, and would require dedicated modeling of the continuum to determine the fractional contribution of the AGN light to the total SED, which is made difficult by the lack of data redder of the F444W band. As such, for the rest of this work we will adopt the $M_{\rm BH}$ derived from the H$\alpha$ luminosity as our final result.

In addition to the mass, we also use the broad H$\alpha$ to derive the bolometric luminosity ($L_{\rm bol}$) of the AGN. We use $L_{\rm bol}=130\times L_{\rm H\alpha}$ \citep{richards06,stern12} and obtain $L_{\rm bol}=(3.46\pm1.10)\times10^{44}$ erg/s. Using the above values, the $L_{\rm bol}/L_{\rm edd}$ is then $\approx0.12$, implying that the object is accreting at a sub-Eddington rate. This is much lower compared to the extreme AGN found at high-$z$ \citep{furtak24,larson23,kokorev23c,maiolino23b}, however some AGN at high-$z$ also appear to be less active \citep{kocevski23,matthee23} or even completely dormant \citep{juodzbalis24}.

\subsection{Ionization}
As briefly stated in the previous sections, some of the nebular narrow lines ratios in the source look elevated when compared to massive star-forming galaxies. In this section we would like to explore this further and investigate the potential ionization mechanisms in the galaxy.

In \autoref{fig:fig5} we show the [\oiii]/H$\beta$ ratio as a function of  [\nii]/H$\alpha$ and [\sii]/H$\alpha$. Given a strong absorption feature present in H$\beta$ (\autoref{fig:fig2}) we only report these ratios as upper limits. Both [\nii] and [\sii] doublets are however detected at S/N$>3$ (\autoref{tab:tab2}) and should provide a more robust diagnostic. Our derived line ratios place us securely in the AGN locus of the BPT \citep{baldwin81} diagram, with regions delimited by \citet{kewley01} and \citet{kauffmann03}. In addition we also derive log$_{10}$([OI]$\lambda6302$/H$\alpha)\sim-0.42$, which again securely places us in the AGN portion of the ionization diagnostic \citep{kewley06}.

We also observe an extremely high ratio between the auroral [\oiii]$\lambda4363$ line and [\oiii]$\lambda5007$ (RO3) of $0.35^{+0.17}_{-0.15}$, and $0.54^{+0.35}_{-0.25}$ when applying the NLR dust correction we derive in \autoref{sec:decr}, respectively. This is quite extreme, and does exceed the dust-corrected RO3=0.32 found in $z\sim8.5$ LRD by \citet{kokorev23c}. The object also appears to be highly ionized when compared to high-$z$ SFGs, with RO3=0.048 presented by \citep{katz23}. Even if our dust-correction is overly aggressive, the observed ratio we find is hard to reconcile with ranges laid out in \citet{nicholls20} and will lead to extreme electron temperatures and densities. 
However, photoionization models presented in \citet{baskin05} alongside RO3 studies from decades of low-$z$ Seyfert investigations \citep[][just to name a few]{koski76,osterbrock78,ferland83,dopita95,nagao01,binette22} do imply that the necessary high temperatures and densities can be reached in the NLR around an AGN. As mentioned in the previous section, high metallicities can cause the [Fe\,\textsc{iii}]$\lambda$4360 line to contaminate the [\oiii]$\lambda$4363 flux \citep{curti17}. Although our fit to the spectrum does not show evidence of [Fe\,\textsc{iii}] emission, the NLR can be easily enriched with metals due to its relatively low mass, resulting in the contribution of this line to [\oiii]$\lambda$4363 to be potentially non-negligible. This could mean that our RO3 is overestimated, but only higher resolution data
from the H-gratings can be used to definitely answer this question. Another possibility for such a high ratio is a highly dense NLR. The critical density of [\oiii]$\lambda$4363 is roughly $50\times$ higher than that of [\oiii]$\lambda$5007, so if the density of some regions within the NLR exceeds 10$^{5.8}$ cm{$^{-3}$}, then RO3 can be artificially boosted as well.


\begin{figure*}
\begin{center}
\includegraphics[width=.99\textwidth]{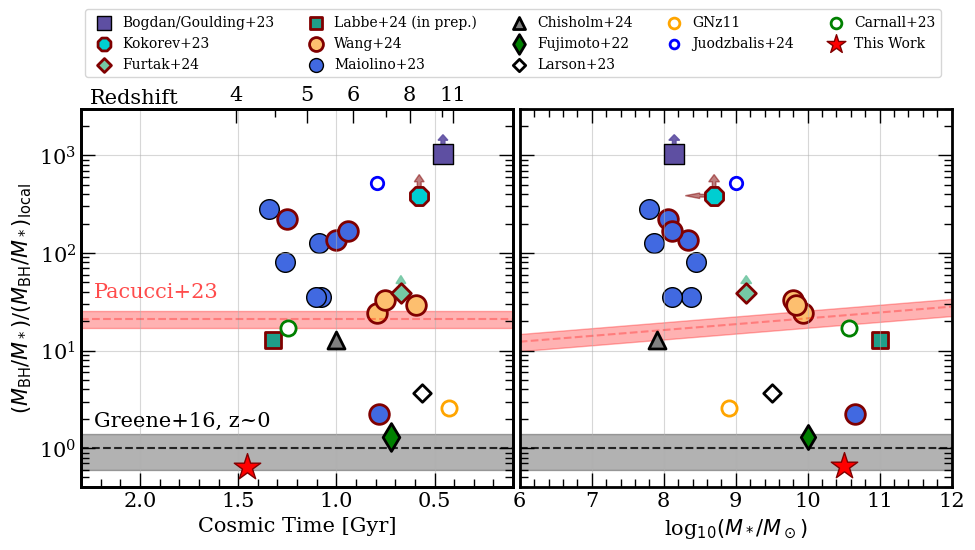}
\caption{Potential evolution of the offset between derived $M_{\rm BH}/M_*$ and a local relation \citep{greene24} across time (left) and stellar mass (right). We show an array of high-$z$ AGN  \citep{bogdan23,goulding23,chisholm24,fujimoto22,furtak24,juodzbalis24,kokorev23c,larson23,
maiolino23,maiolino23b,wang24}. We specifically highlight objects that would be selected as LRDs with a red envelope. In gray and maroon we also show tracks of $M_{\rm BH}/M_*$ relations from \citet{greene24} and \citet{pacucci23}, respectively. We note the emergent downward trend of $M_{\rm BH}/M_*$ toward the local relations.}
\label{fig:fig7}
\end{center}
\end{figure*}

\subsection{AGN Quenching in a Little Red Dot}

With all the pieces in place, we now would like to examine the full picture and reflect on the potential nature of the source. While \srcname\, fulfills color and compactness criteria for LRDs, the colors themselves seem to originate from the evolved stellar population, rather than AGN continuum. All signs point toward \srcname\, being a massive
($\log_{10}(M_*) \sim 10.6$) quiescent galaxy. The SFH inferred from a multitude of absorption lines as well as a strong Balmer break indicate that this galaxy has ceased, at least temporarily, its star formation. In the rest-optical, a combination of broad lines and AGN-like line ratios, betrays a moderately active AGN ($L_{\rm bol}/L_{\rm edd} \sim 0.1$) with a black hole mass of $\log_{10}(M_{\rm BH}) \sim 7.3$, embedded within this massive, quenched galaxy. At the same time, in the rest-UV we detect a strong (EW$\sim400$ \AA) Ly$\alpha$, inconsistent with the low SFR of the host, plus a \mgii\, doublet in emission.
While the evidence for the broad component in \mgii\, is itself tentative, the mere fact that we find this line in emission in a solar metallicity quiescent galaxy, already strongly suggests it originates from an AGN and not star formation \citep{burchett21,witstok21}.

While the LRD we study has clear AGN signatures, such as broad lines and AGN-like ionization, it does appear to have completely settled on the local relations linking black holes and their host galaxies. For example, we find the $M_{\rm BH}/M_*$ in \srcname\, to be completely consistent with the local relations \citep{kormendy13,greene16,greene20}. 
While the narrow lines in the source likely originate from the NLR around an AGN, the availability of clear stellar absorption features allows us to robustly measure the stellar velocity dispersion. We place \srcname\, on the Faber-Jackson sequence \citep{faberjackson} in \autoref{fig:fig6} and find that the object has largely ``settled'' onto the same evolutionary trajectory as local elliptical galaxies that exist on the fundamental plane \citep{djorgovski87}. This is unsurprising as we find that the source does in fact have a lot in common with the massive quiescent galaxies recently uncovered by \textit{JWST} \citep{carnall23,de_graaff24}, rather than the record breaking high-$z$ LRDs. 
In the final section we would like to reflect on how this source potentially came to be, and where it might be headed.

\section{Discussion and Summary} \label{sec:disc_sum}
\subsection{Recently Discovered AGN and Their Hosts}
The last two years of observations with \textit{JWST} have provided us with constraints on $M_{\rm BH}/M_*$ for a large sample of AGN across a wide epoch of early cosmic history ($z=4-9$). We aim to explore these data to determine if there is an evolutionary link between AGN-dominated high-$z$ objects with high $M_{\rm BH}/M_*$ \citep{bogdan23,endsley23,goulding23,kokorev23c,juodzbalis24}, reddened AGNs at $z\sim4$ which have grown their stellar populations significantly over the last $\lesssim1$ Gyr (Labb\'e et al., in prep, \citealt{wang24}), and the earliest quiescent galaxies at $z>4$ \citep{carnall23,de_graaff24}.

In \autoref{fig:fig7}, we show how the $M_{\rm BH}/M_*$ relation evolves for \textit{JWST}-detected AGN and quiescent galaxies containing AGN signatures, compared to the local relation \citep{greene24}, as a function of both cosmic age (redshift) and stellar mass. Starting with the epoch just $\sim400$ Myr ($z\sim11$) after the Big Bang, we cover roughly 1 Gyr of cosmic history in order to reflect on a potential evolutionary path that can tie this picture together.

Formation mechanisms of AGN with elevated $M_{\rm BH}/M_*$ ratios remain elusive. Proposed solutions range from super-Eddington accretion on remnants of Population III stars \citep{furtak24,larson23} and compact starbursts \citep{kroupa20}, to direct collapse seeds \citep{bogdan23,dayal24,goulding23,kokorev23c,natarajan23,maiolino23b} and primordial black holes \citep{LiuBromm2022,dayal24b}. These overmassive black holes could also be explained by by the existence of a new $M_{\rm BH}-M_*$ sequence \citep{pacucci23,pacucci24}, or a high covering fraction, implying that locally-calibrated virial relations which link black hole mass to H$\alpha$ luminosity are not applicable at high-$z$ \citep{maiolino24,pacucci24b}. The perceived excess of these high-mass BHs could also simply stem from a selection bias in luminosity-limited BL samples where the outliers are the only detectable sources and a change to the underlying scaling relations is not necessary \citep{li24}. 
Formation pathway aside, AGN with very little to no detectable host galaxy component have been found across a multitude of extragalactic fields, too many to be ignored.

While the gas in these systems might not be fueling star formation, a considerable amount is still required to sustain AGN accretion.
In their work \citet{maiolino23b} present a sample of high-$z$ AGN (45\% being LRDs, using the same criteria as in \autoref{sec:targsel}) which do not follow the local $M_{\rm BH}-M_*$ relation but align with local dynamical black hole mass relations. This would imply that these systems do contain gas but are inefficient at forming stars. 
The initial view of the far-infrared (FIR) emission in LRDs provided by ALMA reveals a dearth of SF activity, albeit these observations still remain relatively shallow. So far, not a single LRD has been detected in the FIR \citep{akins24,casey24,labbe23}. It has even been suggested that most LRDs are not AGN, but rather compact starburst galaxies instead \citep{perezgonzalez24,williams23b}, as they lack a rising power-law in the mid-infrared (MIR). There is no evidence so far from the MIR/FIR to support or refute that notion, and objects with clear AGN signatures have been found to also show a lack of a dusty torus emission \citep{wang24}, which could be a result of a patchy BLR \citep{maiolino24}.

At lower redshifts ($z\sim4-5$), we are presented with another curiosity: some objects that fulfill the LRD criteria \citep{labbe23,greene24} also appear to contain evolved stellar populations. Model fits to these systems indicate very early and rapid star formation, followed by a significant reduction (\citealt{wang24}; Labb\'e et al., in prep.) or near cessation of SF activity, as in \srcname, as they approach the local $M_{\rm BH}-M_*$ relation.
This source, in particular, provides a unique opportunity to examine the $M_{\rm BH}-M_*$ ratio and the SFH in reddened AGN with unprecedented detail. The clearly stellar-dominated continuum and broad H$\alpha$ emission allow us to extract information concerning both the central black hole and detailed properties of the stellar population in the host galaxy.

Contrasting the SFH of \srcname, with the objects in \citet{wang24}, reveals a scenario where compact, reddened AGN enter a starburst phase $300-400$ Myr before the epoch of observation and then appear to quench. This indicates that there is a stage between elevated and local-like $M_{\rm BH}-M_*$ where these AGN experience rapid star formation, possibly due to a shutdown of AGN feedback, gas virialization, a merger, or a combination of multiple factors. During this stage, we expect that these AGN would be still identifiable (e.g. through high ionization or broad lines), but would also shine brightly in the FIR as they are now actively (and rapidly) forming stars. Given that this phase is short-lived ($\sim100$ Myr), catching these objects in the act is challenging, although it may have already been accomplished.

In their pre-\textit{JWST} work, \citet{fujimoto22} presented GNz7q, a compact object at $z\sim7.2$, which potentially bridges quasars at cosmic dawn with galaxies at cosmic noon. The spectral shape and extremely high UV-luminosity surface density 
of GNz7q indicate it cannot be explained by extreme star formation alone, suggesting its AGN nature. At the same time it is embedded in an extremely FIR luminous host starburst galaxy
(SFR$_{\rm IR}>1600$ $M_\odot$/yr), all within $\sim 480$ pc, 
precisely the type of transitioning, short-lived object we were looking for. Moreover, GNz7q is extremely faint in X-rays, suggesting a Compton-thick, super-Eddington black hole accretion disk, a characteristic observed in high-$z$ AGN \citep{endsley23,furtak24,greene24,kokorev23c,maiolino23}. It does appear that star-formation can take place in galaxies hosting massive AGN and can do so at quite an accelerated rate. Notably, GNz7q is a very extreme example but there likely exists a small population of LRD hosts in the starburst phase, that is yet to be identified. It is very possible that these are NIRCam dark,  due to extreme dust obscuration ($A_{\rm V}>4-5$) and low surface brightness, making the detection of rest-UV/optical emission extremely challenging. The existence of such dusty systems at high redshift was briefly explored in \citet{kokorev23a}, and recently, a NIRCam-dark source was identified by \citet{perezgonzalez24b}, yet the exact number densities of these objects are still unclear.

\subsection{Evolution and Link to the Quiescent Galaxies}
A multitude of studies preceding the launch of \textit{JWST} have laid out the possibility that cores of the most massive galaxies in our local Universe have, in fact, formed very rapidly. This is generally thought to happen via a burst of star-formation at $z>5-6$, followed by a rapid quenching \citep{thomas05,conroy14,vandokkum15}.

The identification and detailed analysis of $z=3-5$ quiescent galaxies, made possible with NIRSpec, have further supported the notion of rapid star-formation and subsequent quenching \citep{carnall23,de_graaff24,glazebrook24}. Some of these galaxies seemingly retain their AGN signatures as either broad lines \citep{carnall23} or high ionization \citep{de_graaff24}. The sample presented recently in \citet{wang24} seems to show the progenitors of some of these quiescent galaxies as their star-formation undergo a truncation. The authors propose a scenario where massive galaxies have simultaneously grown alongside their SMBHs and then proceed to quench rapidly, however we would like to offer an alternative interpretation. 

It is possible that the systems with elevated $M_{\rm BH}/M_*$ we find with \textit{JWST} will undergo episodes of intense star formation. These are likely induced through merging, as many LRDs were found to exist in overdensities (\citealt{fujimoto23_uncover,greene24,kokorev23c}; Labb\'e et al., in prep.). Following a short-lived yet intense starburst, like the one observed in \citet{fujimoto22} or predicted by \citet{carnall23} and \citet{de_graaff24}, the stellar mass of the host galaxy catches up to its over-massive BH and settles onto the local BH-host relations. The next likely step of their evolution is further expansion and transition toward the giant ellipticals we see locally. 

Having undergone a burst and then a rapid quenching (\autoref{fig:fig4}),  \srcname\, appears to agree with this picture. Dominated by stellar light the object still contains clear AGN signatures, which suggests that rapid quenching could have been driven by AGN feedback rather than gas depletion. We propose that while some high-$z$ AGN may not initially follow concurrent growth of BHs and their stellar populations, they will eventually ``settle'' onto the local relations, as time progresses. 

Generally, elliptical galaxies are thought to form through processes such as mergers of smaller galaxies. During these mergers, the system undergoes violent relaxation, redistributing energy and leading to a state of equilibrium. The duration of such a process will be longer for extended and less massive systems, and shorter for the most massive and compact ones. We can see how close the source is to a "settled" virialized state by examining its position on the  
Faber-Jackson relation \autoref{fig:fig6}. We find that \srcname\, occupies a position very similar to the local ellipticals, implying that such virialization may already take place quite early, at $z\sim4$, potentially for the most massive objects. On the other hand BHs with measured $\sigma$ from \citet{maiolino23b}, especially the ones with similar stellar mass, are at higher redshift compared to \srcname\, and are likely still evolving into their equilibrium state. The achievement of a virial equilibrium is crucial to our argument, as then it would make it difficult for galaxies to have significant $M_{\rm BH}/M_*$ variations as their stellar masses exceed $\sim10^{10}$ M$_\odot$, ultimately aligning them with local relations.

Despite this, we find that our source is offset from the local $M_{\rm BH}-\sigma$ relation \citep[e.g.][]{kormendy13,greene20}. Our derived $\sigma_*$ suggests a $M_{\rm BH}\sim10^9$ $M_\odot$, $\sim2$ dex higher than our current measurement. It is therefore possible that the stellar component of \srcname\ still requires more time to relax and expand from its very compact configuration to align with the local $M_{\rm BH}-\sigma$ trend, due to the conservation of angular momentum, driving down the $\sigma$. Importantly, we do not observe any compact objects with LRD colors below $z\sim4$ \citep{kocevski24}, supporting the idea that these objects have begun to expand and dynamically relax into structures resembling lower-$z$ ellipticals. Furthermore, studies of elliptical galaxies from $z=2$ to 0 show that their effective radii grow by a factor of $2-4$ over $\sim10$ Gyr timescales \citep{vanderwel08,cassata11,vanderwel14,xie15}. This observed growth and relaxation process is consistent with the transformation of LRDs into the quiescent elliptical galaxies we see at lower redshifts.

It is worth noting that objects presented in \autoref{fig:fig7} do not necessarily follow these exact evolutionary paths. A large fraction of high-$z$ AGN have poorly measured $M_*$ as the AGN continuum dominates the emission making it difficult to separate the two \citep[see discussion in][]{greene24,wang24}. Dynamical masses can be employed instead, however the same problem manifests itself when trying to decide whether the narrow nebular lines originate from the NLR or the host galaxy itself. In the case of \srcname, however, the stellar-dominated continuum and the presence of absorption lines allows for robust measurements of both $M_*$ and $M_{\rm dyn}$.

\subsection{Final Remarks}
Using the NIRSpec PRISM and medium resolution G140M, G235M, G395M observations of the GOODS-N field \citep{deugenio24}, we present the investigation of a little red dot that shows signs of both an evolved stellar population and AGN activity at $z=4.13$. A careful examination of the spectrum of the source shows a significant ($>4\sigma$) BL component present in the H$\alpha$ line with a FWHM of $\sim2500$ km/s. As we do not observe similar broad features in the strongly detected [\nii], [\sii] or the [\oiii] lines, we suggest that the broadening exhibited by H$\alpha$ is a sign of an actively accreting black hole, rather than large-scale outflows. A further evidence for the AGN activity are high ionization inferred from the BPT diagram, the RO3 line diagnostic, as well as a strong Ly$\alpha$ emission with an EW$_0\sim400$ \AA. In the rest-UV we also report a tentative broad component in the \mgii\, doublet, which could further strengthen our AGN interpretation.

Despite being compact and showing clear AGN signatures, \srcname\, appears to be different from the high-$z$ LRDs explored with NIRSpec \citep{furtak24,kokorev23c,greene24}. We detect a massive (log$_{10}(M_*)\sim10.6$) and evolved host, which dominates the continuum of our object, as evidenced by a strong Balmer break as well as a multitude of absorption lines. Detailed SFH fitting reveals that our object has formed most of its stellar population in a short ($100-200$ Myr) burst, and then quenched rapidly roughly 200 Myr ago. It is worth noting that stellar masses of LRDs are generally poorly constrained due to difficulties separating the AGN and stellar continuum. Despite that, a strong stellar that we detect in \srcname\, would be clearly visible even at high-$z$, provided it was there.

We contrast the position of \srcname\, on the $M_{\rm BH}/M_*$ vs $z$ and $M_*$ plane to all other high-$z$ LRDs and BL AGN uncovered with \textit{JWST}. This examination reveals a potential evolutionary path that starts with overmassive BH in the early Universe, which then form stars in a rapid burst and slowly descend onto the local $M_{\rm BH}-M_*$ relation occupied by local ellipticals.

Finally, we detect very strong Ly$\alpha$ emission, inconsistent to have come solely from star formation, and a tentative broad line in \mgii\, doublet. This would mean that, for the first time, we now see clear AGN signatures in the rest-UV spectra of LRDs. Despite that we can not yet ascertain the full origins of blue light in LRDs. 
While we detect emission lines with strong AGN signatures in the rest-UV, we also find it to be marginally resolved ($\sim 500$ pc), too large to have come just from the central AGN alone. It is likely that some AGN light indeed escaped the thick dust cover through either scattering and some patchy coverage, but the continuum itself is likely to be stellar in origin. Rest-UV emission in \srcname\, appears to originate both from stars and AGN, however this still might not be true for the entire population, especially as we move to higher redshifts where AGN seem to be much more dominant. Although it is worth noting still, that observational bias can drive the high incidence of objects with elevated $M_{\rm BH}/M_*$ \citep[e.g.][]{li24}. Only dedicated deep studies of the rest-UV emission of LRDs with medium and high resolution NIRSpec gratings onboard \textit{JWST} can shine more light on this mystery.

\acknowledgements
VK acknowledges support from the University of Texas at Austin Cosmic Frontier Center. We thank the JADES team for providing the scientific community with these beautiful \textit{JWST} data.
This work is based on observations made with the NASA/ESA/CSA \textit{James Webb Space Telescope}. The data were obtained from the Mikulski Archive for Space Telescopes at the Space Telescope Science Institute, which is operated by the Association of Universities for Research in Astronomy, Inc., under NASA contract NAS 5-03127 for \textit{JWST}.
Some of the data products presented herein were retrieved from the Dawn \textit{JWST} Archive (DJA). DJA is an initiative of the Cosmic Dawn Center, which is funded by the Danish National Research Foundation under grant No. 140. The X-ray datasets were obtained from the Chandra Xray Center.

\software{BAGPIPES \citep{carnall18,carnall19}, EAZY \citep{brammer08}, grizli \citep{grizli}, msaexp \citep{msaexp}.} 

\facilities{\jwst}


\bibliographystyle{aasjournal}
\bibliography{refs}

\end{document}